\documentclass{mem}
\usepackage{natbib}\usepackage{txfonts}\usepackage{balance}
\usepackage{graphicx}
\usepackage[a4paper]{hyperref}
\idline{XX}{1}
\begin{document}
\def\teff{$T\rm_{eff }$}
\def\kms{$\mathrm {km s}^{-1}$}
\def\ltsim{\raisebox{-.5ex}{$\;\stackrel{<}{\sim}\;$}}
\def\gtsim{\raisebox{-.5ex}{$\;\stackrel{>}{\sim}\;$}}

\title{
Shock-in-jet model for quasars and microquasars
}

   \subtitle{}

\author{
Marc T\"urler
          }

  \offprints{M. T\"urler}

\institute{
ISDC, Geneva Observatory, University of Geneva, ch. d'Ecogia 16, 1290 Versoix, Switzerland
\email{marc.turler@unige.ch}
}



\abstract{
We present the theoretical background and detailed equations for the synchrotron emission of a shock wave propagating in a relativistic jet. We then show how the evolution of an outburst in this shock-in-jet scenario can be analytically described and parameterized to be fitted to multi-frequency lightcurves of galactic and extragalactic sources. This is done here for the first time with a completely physical description of the jet and the shocked gas, while previous studies used a more phenomenological approach based on the observed properties of the outbursts. Another interesting addition to previous work is the introduction of a low-energy cut of the electron energy distribution that allows for much more diverse synchrotron spectral shapes. To demonstrate and illustrate the new methodology, we present results of infrared-to-radio lightcurve fitting of a succession of outbursts observed in 1994 in the microquasar Cyg X-3. We find that the diversity of outbursts in shape, amplitude, frequency range and timescale can be fairly described by varying only the strength of the shock and its build-up distance from the apex of the jet. A rapid build-up results in high-frequency outbursts evolving on short timescales, while slowly evolving, low-frequency outbursts form and evolve further out in the jet. We conclude by outlining future developments, in particular the inclusion of the associated synchrotron self-Compton emission at X-rays and gamma-rays.
}
\maketitle{}

\section{Introduction}

25 years ago, \citet{MG85} introduced a shock-in-jet model to explain a giant flare in the quasar 3C~273 observed in 1983 \citep{RGC83}. The observations showed a rise of the flux density of the synchrotron self-absorption turnover together with a slight move of the turnover from high- to low-frequencies. This behavior was not compatible with the plasmon model of \citet{L66} describing the adiabatic expansion of a spheroidal blob of plasma. It was, however, compatible with a shock model taking into account Compton cooling of the electrons in the initial stage of its evolution \citep{MG85}. Until now, this model could not be disproved and was found to give a good description of the observations of both galactic and extragalactic sources of relativistic jets \citep[3C~273:][]{TCP99,TCP00} \citep[3C~279:][]{LTV06} \citep[GRS~1915+105:][]{TCC04} \citep[Cyg~X-3:][]{LTH07,MRT09}.

Based on the experience gained on the modeling of the flaring behavior of these sources, we describe here a fully physical approach that we intent to use in the future. It has the advantage to fit directly the physical jet properties (electron energy distribution, magnetic field, etc.) rather than the observables (frequency and flux density at the spectral turnover, etc.). This change will also allow us to soon incorporate the associated synchrotron self-Compton component in a self-consistent way.
We present the main equations of the shock model starting from standard synchrotron theory in  Sect.~\ref{turler:theory} and the method of shock-in-jet modeling in Sect.~\ref{turler:method}. We then present results obtained for Cyg X-3 in Sect.~\ref{turler:results} and conclude with future perspectives in Sect.~\ref{turler:conclusion}.

\section{Theory}
\label{turler:theory}
Synchrotron theory as derived by \citet{P70}, for instance, is relatively complex. Let's assume a homogeneous synchrotron source with a powerlaw electron energy/impulsion distribution given by $n(\gamma)=K\,\gamma^{-p}$ for $\gamma_{\mathrm{min}}\!\le\!\gamma\!\le\!\gamma_{\mathrm{max}}$, where $n$ is the electron number density in cm$^{-3}$ and $\gamma$ is the Lorentz factor  of the relativistic electrons, or more precisely the product $\beta\gamma$ for mildly relativistic electrons. This results in equations for the emission $\varepsilon_{\nu}$ and absorption $\kappa_{\nu}$ coefficients that have their simplest form when expressed as a function of the cyclotron frequency $\nu_B\equiv eB/(2\pi mc)$ where $e$ and $m$ are the charge and mass of the electron and $B$ is the component orthogonal to the line of sight of an uniform magnetic field in the source. The equations are:
\begin{eqnarray}
\label{turler:epsilon_nu}
\varepsilon_{\nu}&=&\frac{e^2}{8c}\,g_{\varepsilon}(p)\,K\,\nu_B^{(p+1)/2}\,\nu^{-(p-1)/2}\\
\label{turler:kappa_nu}
\kappa_{\nu}       &=&\frac{e^2}{16mc}\,g_{\kappa}(p)\,K\,\nu_B^{(p+2)/2}\,\nu^{-(p+4)/2}
\end{eqnarray}
where $g_{\varepsilon}(p)$ and $g_{\kappa}(p)$ are a product of Euler gamma functions, $\Gamma$, given by:
\begin{eqnarray}
g_{\varepsilon}(p)\!&\!=\!&\!3^{\frac{p}{2}}\left(\frac{p\!+\!\frac{7}{3}}{p+1}\right)~\Gamma\left(\frac{3p\!-\!1}{12}\right)~\Gamma\left(\frac{3p\!+\!7}{12}\right)\\
g_{\kappa}(p)       \!&\!=\!&\!3^{\frac{p+1}{2}}\!\left(p\!+\!\frac{10}{3}\right)\,\Gamma\left(\frac{3p\!+\!2}{12}\right)\,\Gamma\left(\frac{3p\!+\!10}{12}\right)
\end{eqnarray}
The latter functions are slowing varying in the range of interest and have both a minimum at $g_{\varepsilon}(p\!\approx\!2.6)\approx8.2$ and at  $g_{\kappa}(p\!\approx\!0.9)\approx26.4$.

Solving the differential equation of the radiative transfer then gives the specific intensity $I_{\nu}$ in erg\,s$^{-1}$\,cm$^{-2}$\,Hz$^{-1}$\,sr$^{-1}$ as:
\begin{equation}
\label{turler:I_nu}
I_{\nu}\!=\!\frac{\varepsilon_{\nu}}{\kappa_{\nu}}(1-\mbox{e}^{-\tau_{\nu}})\!=\!2m\frac{g_{\varepsilon}(p)}{g_{\kappa}(p)}\frac{\nu^{5/2}}{\nu_B^{1/2}}(1-\mbox{e}^{-\tau_{\nu}}).
\end{equation}
One recognizes the typical synchrotron spectrum with an optically thick part ($\tau_{\nu}\!\gg\!1$) going as $\nu^{5/2}$ and the optically thin part ($\tau_{\nu}\!\ll\!1$) following a $\nu^{-(p-1)/2}$ dependence since $I_{\nu}^{\mathrm{thin}}=x\,\varepsilon_{\nu}$, which is obtained by replacing ($1-\mbox{e}^{-\tau_{\nu}}$) by $\tau_{\nu}=x\,\kappa_{\nu}$ in Eq.~(\ref{turler:I_nu}), where $x$ is the thickness of the source along the line of sight.
We note that this standard synchrotron spectrum strictly holds only if there is no significant radiative cooling of the electrons and if $\nu_{\mathrm{min}}$ tends to 0 and $\nu_{\mathrm{max}}$ tends to infinity, where $\nu_{\mathrm{min}}\!=\!\nu_B\gamma_{\mathrm{min}}^2$ and $\nu_{\mathrm{max}}\!=\!\nu_B\gamma_{\mathrm{max}}^2$. A synchrotron spectrum can have a much different shape if radiative cooling is important or if $\nu_{\mathrm{min}}$ happens to be higher than the frequency of synchrotron self-absorption, $\nu_{\mathrm{abs}}$, as illustrated by e.g. \citet[Fig.~1]{GS02}. In particular, in the slow cooling case, the optically thin spectral slope will break to a $\nu^{1/3}$ dependence at frequencies between $\nu_{\mathrm{abs}}$ and $\nu_{\mathrm{min}}$.

\begin{figure*}[t!]
\resizebox{\hsize}{!}{\includegraphics[clip=true]{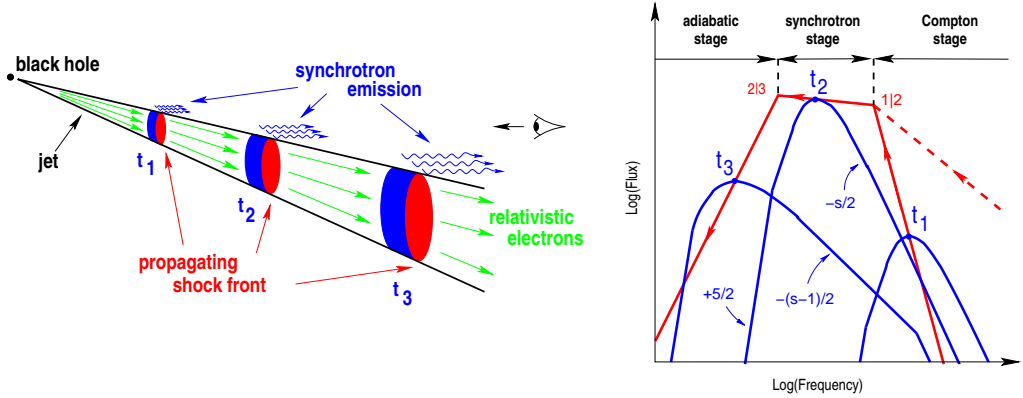}}
\caption{\footnotesize
Schematic representation of a propagating shock wave in a relativistic jet and the three-stage evolution of the associated synchrotron outburst according to the model of \citet{MG85} and with the modification of the Compton stage proposed by \citet{BA00} (dashed line). Here $s\equiv p$.
}
\label{turler:f1}
\end{figure*}

For the emission of a shock wave in a quasar, we consider a source located at a luminosity distance $D_L$ -- corresponding to a redshift of $z$ -- with a slab geometry of thickness $x$ and radius $R$, corresponding to the half-width of the jet (see Fig.~\ref{turler:f1}). The source is moving with a speed $\beta c$ corresponding to a bulk Lorentz factor of $\Gamma=(1-\beta^2)^{-1/2}$ in a direction making a small angle $\theta$ with the line-of-sight, which results in a Doppler boosting factor $\delta=\Gamma^{-1}(1-\beta\cos{\theta})^{-1}$. The optically thin flux density $F_{\nu}^{\mathrm{thin}}=\Omega\,I_{\nu}^{\mathrm{thin}}=\Omega\,x\,\varepsilon_{\nu}$ of a uniform source sustaining a solid angle $\Omega=\pi R^2 (1\!+\!z)^4/D_L^2$ is then given by:
\begin{equation}
\label{turler:F_nu}
F_{\nu}^{\mathrm{thin}}\!=\!\frac{\pi e^2}{8c}g_{\varepsilon}(p)\frac{\delta^{(p+5)/2}}{(1\!+\!z)^{(p-3)/2}}\frac{R^2}{D_L^2} x K \frac{\nu_B^{(p+1)/2}}{\nu^{(p-1)/2}}
\end{equation}
where $x$, $R$, $K$ and $B$ are expressed in the frame of the moving source, whereas $\nu$ and $F_{\nu}$ are the observed quantities.
We can also derive the frequency of synchrotron self-absorption by setting $\tau_{\nu}(\nu_{\mathrm{abs}})=x_{\mathrm{abs}}\,\kappa_{\nu}(\nu_{\mathrm{abs}})=1$, which gives:
\begin{equation}
\label{turler:nu_abs}
\nu_{\mathrm{abs}}=\frac{\delta}{1\!+\!z} \left(\frac{e^2g_{\kappa}(p)}{16mc} x_{\mathrm{abs}}\,K\,\nu_B^{(p+2)/2}\right)^{2/(p+4)}
\end{equation}

The two last equations expressed as proportionalities are the starting point of the shock model of \citet{MG85}. The critical parameter that defines the well-known three-stage evolution of the outburst (see Fig.~\ref{turler:f1}) is the thickness $x$ of the emission region along the line of sight. The evolution of a spherical plasmon expanding adiabatically in 3-dimensions is obtained by replacing $x$ by $R$ and by letting $K$ evolve with $R^{-(p+2)}$ and $B$ with $R^{-2}$ \citep{L66}. The result is a decrease of flux density, while the spectral turnover is moving towards lower frequencies. \citet{MG85} assume the source to be a cylinder of thickness $x_R=f_R R$ -- a constant fraction $f_R$ of $R$ -- in a conical jet expanding in the two directions perpendicular to the jet axis such that $K\propto R^{-2(p+2)/3}$ and with the magnetic field to be preferentially orthogonal to the jet axis ($B\propto R^{-1}$) which leads to a slightly shallower decrease of the flux density, but the trend is the same. To explain the rise in flux density observed in the early phases of the the 1983 outburst in 3C~273, \citet{MG85} had the idea to strongly limit the source volume at the onset of the outburst. This is achieved by considering that $x\ll x_R$ at the beginning of the shock evolution because the thickness is limited by the distance electrons can travel from the shock front before they cool down through radiative losses, such that $x(\nu)=\beta_{\mathrm{rel}} c t_{\mathrm{cool}}(\nu)$. \citet{BA00} propose a modification of the expression of $x$ used by \citet{MG85} as:
\begin{equation}
\label{turler:x_nu}
x(\nu)=\frac{3mc^2\beta_{\mathrm{rel}}}{8\,\sigma_{\mathrm{T}}}\frac{\nu_B^{1/2}}{U_B+U_{\mathrm{S}}}\left(\frac{\delta}{1\!+\!z}\right)^{1/2}\!\nu^{-1/2}
\end{equation}
where $\sigma_{\mathrm{T}}$ is the Thomson cross-section and $\beta_{\mathrm{rel}}$ is the average speed -- in units of $c$ -- of the electrons relative to the shock front, which takes a value of $0.3$ for a relativistic shock \citep{B10}\footnote{The factor $3/8$ in Eq.~(\ref{turler:x_nu}) corresponds roughly to the cooling time needed for the final (cooled) emission frequency to be at half the initial frequency,  $\nu_{\mathrm{f}}/\nu_{\mathrm{i}}\!=\!(\gamma_{\mathrm{f}}/\gamma_{\mathrm{i}})^2\!=\!1/2$, rather than $\gamma_{\mathrm{f}}\!=\!\gamma_{\mathrm{i}}/2$, which corresponds to a factor $3/4$ (Bj\"ornsson~C.-I., private comm.).}.
The dominant cooling can either be inverse-Compton on the virtual photons of the magnetic field with an energy density $U_B=B^2/(8\pi)$ (synchrotron cooling) or on the actual synchrotron photon with an energy density of $U_{\mathrm{S}} = (4\pi/c)\int_{\nu}I_{\nu}\,d\nu$ (first-order Compton cooling). In any case, this radiative cooling results in a steeper optically thin slope following a $\nu^{-p/2}$ dependence compared to the adiabatic cooling, as can be seen by inserting Eq.~(\ref{turler:x_nu}) into Eq.~(\ref{turler:F_nu}). But even in the adiabatic stage, a cooling break remains at $\nu_{\mathrm{cool}}$ -- defined by $x(\nu_{\mathrm{cool}})\equiv x_R$ -- resulting in a spectral index of $-p/2$ at higher frequencies.

The next step is to insert Eq. ~(\ref{turler:x_nu}) for $x_{\mathrm{abs}}\equiv x(\nu_{\mathrm{abs}})$ into Eq.~(\ref{turler:nu_abs}) to get:
\begin{equation}
\label{turler:nu_abs2}
\nu_{\mathrm{abs}}^{\prime\,(p+5)/2}\!=\!\frac{3e^2c\beta_{\mathrm{rel}}g_{\kappa}(p)}{128\,\sigma_{\mathrm{T}}\,(U_B\!+\!U_{\mathrm{S}})}\,K\,\nu_B^{(p+3)/2}\!,
\end{equation}
where we introduce $\nu^{\prime}\!=\!\nu\,(1\!+\!z)/\delta$ to denote the frequency measured in the frame of the moving source.

The main difficulty comes now with the evaluation of $U_{\mathrm{S}}$. The main contribution to $U_{\mathrm{S}}$ comes from the optically thin part of the emission $I_{\nu}^{\mathrm{thin}}=x(\nu)\,\varepsilon_{\nu}$, so that we can calculate $U_{\mathrm{S}}$ by integrating the optically thin spectrum extending between $\nu_1$ and $\nu_2$ by using Eqs.~(\ref{turler:epsilon_nu}) and (\ref{turler:x_nu}), as:
\begin{equation}
\label{turler:U_S}
U_{\mathrm{S}}^2\!+\!U_B U_{\mathrm{S}}\!=\!\frac{3\pi me^2\beta_{\mathrm{rel}}}{16\sigma_{\mathrm{T}}g_{\varepsilon}^{-1}(p)}K\nu_B^{\frac{p+2}{2}} \!\!\! \int_{\nu_1^{\prime}}^{\nu_2^{\prime}} \!\!\!\! \nu^{\prime-p/2}\!d\nu^{\prime}\!\!
\end{equation}
For $p\!\neq\!2$, the integral in Eq.~(\ref{turler:U_S}) is equal to $2(\nu_2^{\prime(2-p)/2}-\nu_1^{\prime(2-p)/2})/(2\!-\!p)$ and then the second order equation can be solved for $U_{\mathrm{S}}$. While $\nu_2$ is always $\nu_{\mathrm{max}}$, $\nu_1$ is the highest of the two frequencies $\nu_{\mathrm{abs}}$ or $\nu_{\mathrm{min}}$. Whether this is the case is however not known a priori since $\nu_{\mathrm{abs}}$ depends on $U_{\mathrm{S}}$ in Eq.~(\ref{turler:nu_abs2}). The way to proceed is to calculate $U_{\mathrm{S}}$ with $\nu_1$ set to $\nu_{\mathrm{min}}$ to get a first guess of $\nu_{\mathrm{abs}}$, which we call $\nu_{\mathrm{abs},1}$. Even if it turns out that $\nu_{\mathrm{abs},1}>\nu_{\mathrm{min}}$, we can still stay with this result for $U_{\mathrm{S}}$ in most cases. It is only if in addition $p>2$ and $U_{\mathrm{S}}$ is comparable or greater than $U_B$ that taking $\nu_{\mathrm{abs}}$, instead of $\nu_{\mathrm{min}}$, for $\nu_1$ would have a significant effect. In the latter case, the following expression derived from Eqs.~(\ref{turler:nu_abs2}) and (\ref{turler:U_S}) was found to be a good approximation for the right-hand side of Eq.~(\ref{turler:U_S}) when $p\gtsim2$:
\begin{equation}
\label{turler:U_S2}
\frac{3\pi me^2\beta_{\mathrm{rel}}\,K\,\nu_B^2}{8(p\!-\!2)\,\sigma_{\mathrm{T}}g^{-1}_{\varepsilon}(p)}\left(1-\left(\frac{\nu_{\mathrm{abs},1}}{\nu_{\mathrm{max}}}\right)^{(p-2)/2}\right).
\end{equation}

Once $U_{\mathrm{S}}$ is derived, it is simple to calculate $x_{\mathrm{abs}}$ from a relation obtained by inserting Eq.~(\ref{turler:nu_abs}) into Eq.~(\ref{turler:x_nu}).
We then limit $x_{\mathrm{abs}}$ to its maximum value of $x_R$ applying to the final decay stage and calculate $\nu_{\mathrm{abs}}$ through Eq.~(\ref{turler:nu_abs}) and $F_{\mathrm{abs}}\equiv F_{\nu}^{\mathrm{thin}}(\nu_{\mathrm{abs}})$ through Eq.~(\ref{turler:F_nu}).

\section{Method}
\label{turler:method}
Applying the theory outlined in Sect.~\ref{turler:theory} to multi-wavelength observations of quasars and microquasars is the next challenge. For this, we need to define the temporal evolution of the physical parameters $K$ and $B$; of the spectral turnover defined by $\nu_{\mathrm{abs}}$ and $F_{\mathrm{abs}}$; and of the frequency of additional spectral breaks at $\nu_{\mathrm{min}}$, $\nu_{\mathrm{max}}$, and $\nu_{\mathrm{cool}}$.

We consider a jet that is not necessarily conical, but with an opening radius $R$ increasing as a powerlaw of the distance along the jet $X$ as $R=R_0(X/X_0)^r$, where $X_0$ is an arbitrary normalization length. If $r<1$ we get a collimating jet, whereas if $r>1$ we get a trumpet-shaped jet with decreasing collimation.

We can then assume powerlaw dependences of the electron number density $K\!=\!K_0(R/R_0)^{-k}$ and of the magnetic field $B\!=\!B_0(R/R_0)^{-b}$ along the undisturbed jet. For an adiabatic jet flow expanding in two dimensions perpendicular to the jet axis, we have $k_{\mathrm{ad}}\!=\!2(p+2)/3$, which can be seen as a lower-limit if there are also significant radiative losses. The value of $b$ shall be between $1$ and $2$ corresponding, respectively, to the perpendicular, $B_{\perp}$, and parallel, $B_{||}$, components of the field. There are however two arguments to prefer a value of $b\!=\!4/3$. First by assuming a turbulent magnetic field one gets $B_{\perp}\!=\!2B_{||}$ -- rather than $B_{\perp}\!=\!B_{||}$ -- that corresponds to $b\!=\!4/3$, simply because there are two dimensions of space perpendicular to the jet axis and only one parallel to it. Secondly, by assuming equipartition between the energy densities of the electrons, $U_{\mathrm{e}}\!=\!mc^2\int n(\gamma) \gamma d\gamma$, and the magnetic field, $U_B\!=\!B^2/(8\pi)$, one gets $B^2/K\propto R^{2(p-2)/3}$ and thus $b_{\mathrm{eq}}\!=\!(k/2)\!-\!(p\!-\!2)/3$, which is equal to $4/3$ for $k\!=\!k_{\mathrm{ad}}$.

We note that in the shock-in-jet scenario of \citet{MG85}, the electrons are already accelerated in the undisturbed jet and the shock wave mainly compresses the flow locally without strongly accelerating particles.  The strength of the compression factor, $\eta$ -- defined as the ratio of electron number densities in front of and behind the shock front -- is a good candidate to account for differences from one outburst to the next. We therefore leave $\eta_i$ as a free parameter for each outburst $i$. The evolution of $B$ and $K$ for outburst $i$ is thus given by $B_i\!=\!\eta_i B_0(R/R_0)^{-b}$ and $K_i=\eta_i K_0(R/R_0)^{-k}$. The energy gain of each electron as it crosses the shock front, $\xi$, has a minimal value of $\xi_{\mathrm{ad}}\!=\!\eta^{1/3}$ in the case of an adiabatic acceleration process \citep{MG85}. By taking this as the baseline, we get $\gamma_{\mathrm{m}}\!=\!\eta_i^{1/3}\gamma_{\mathrm{m},0}(R/R_0)^{-(k-2)/(p-1)}$, where the subscript `m' stands for `min' or `max' and where the exponent of $R$ -- equal to $-2/3$ for $k\!=\!k_{\mathrm{ad}}$ --  is deduced for a two-dimensional expansion, i.e. with the electron number density decreasing as $n\!=\!\int n_{\gamma}d\gamma\!\propto\!R^{-2}$. This leads to $\nu_{\mathrm{m}}\!=\!(\delta/(1\!+\!z))\,\nu_B\gamma_{\mathrm{m}}^2\!\propto\!\eta_i^{2/3}R^{-b-2((k-2)/(p-1))}$, where we assumed a constant Doppler factor $\delta$.
The evolution with $R$ of the two final spectral breaks,  $\nu_{\mathrm{cool}}$ and $\nu_{\mathrm{abs}}$, is more complex as they depend on $U_\mathrm{S}$, which can be obtained from Eq.~(\ref{turler:U_S}).

As we are interested in the evolution of an outburst with time rather than with $R$ or $X$ we define an origin of time for the jet flow at the apex of the jet, $t_{\mathrm{jet}}(X\!=\!0)\!=\!0$. In the frame of the black hole, we have $X\!=\!\beta\,c\,t_{\mathrm{jet}}^{\mathrm{BH}}$, which gives $X\!=\!\beta\,c\,\Gamma\,\delta\,t_{\mathrm{jet}}/(1\!+\!z)$, by expressing the time in the observer's frame. To account for a possible acceleration or deceleration of the jet flow, the simplest is to express it as a powerlaw dependence on $X$, as $\beta\,\Gamma\!=\!\beta_0\Gamma_0(X/X_0)^g$ and $\delta\!=\!\delta_0(X/X_0)^d$, where we linked $\beta$ and $\Gamma$ to account for both mildly and strongly relativistic jets and we separated the Doppler factor to further account for the emission of a curved jet. For a straight jet, we have qualitatively that $d\!\approx\!0$ for $\beta\,\Gamma\!<\!1$, whereas $d$ depends on $g$ for $\beta\,\Gamma\!>\!1$ as $d\!\approx\!g$ for $\theta<1/\Gamma$ and as $d\!\approx\!-g$ for $\theta>1/\Gamma$. We can then integrate  the expression of $dt_{\mathrm{jet}}\!=\!(1\!+\!z)/(c\,\beta\,\Gamma\,\delta)dX$ from zero to $t_{\mathrm{jet}}$ to get:
\begin{equation}
\label{turler:R}
R\!=\!R_0\left(\frac{X}{X_0}\right)^r\!\!=\!R_0\left(\frac{t_{\mathrm{jet}}}{t_{\mathrm{jet},0}}\right)^{r/(1-g-d)},
\end{equation}
where $t_{\mathrm{jet},0}\!\equiv\!(1\!+\!z)X_0/(c\,\beta_0\Gamma_0\delta_0(1\!\!-\!\!g\!\!-\!\!d))$. This relation links $R$ to the observable $t_{\mathrm{jet}}\!=\!t\!-t_{X=0}$, where $t_{X=0}$ can be fitted for each outburst present in the observed lightcurves. It is however more natural to fit the time of the peak of the outburst, $t_{\mathrm{p},i}$, and the corresponding distance along the jet, $X_{\mathrm{p},i}$, for each flare $i$ and to use this to calculate $t_{X=0,i}$. The choice of $X_{\mathrm{p}}$ as a free parameter for any outburst in addition to $\eta_i$ is motivated by the finding that the typical distance where the shock evolves seems to be the main driver of the different properties of individual outbursts found in Cyg~X-3 \citep{MRT09}. As it is not physical to set the compression $\eta$ to zero until reaching $X_{\mathrm{p},i}$ and then abruptly let it jump to its final value $\eta_i$, we assume a linear increase of $\eta$ with $X$ from the apex of the jet to $X_{\mathrm{p},i}$.

To summarize, the parameters we fit to a set of multi-frequency lightcurves are some of the indices $p$, $r$, $k$, $b$, $g$, and $d$ depending on the assumptions, as well as some of the normalizations $K_0$, $B_0$, $\beta_0\Gamma_0$, $\gamma_{\mathrm{min},0}$,  $\gamma_{\mathrm{max},0}$, and the factor $f_R$. With this and the fixed value of $X_0$, we can get $R_0\!=\!X_0\tan{\phi_0}$ based on the jet opening half-angle $\phi_0\!\approx\!10\degr/\Gamma_0$ deduced by \citet{JML05}. $\delta_0$ also derives from $\Gamma_0$ by using the jet angle to the line-of-sight, $\theta$, taken from studies of superluminal motion in the considered sources. We relate the observed time in the lightcurve $t$ to the radius of the jet $R$ through Eq.~(\ref{turler:R}) and we can thus calculate the evolution with observed time of $K$, $B$ and $x_{\mathrm{abs}}$ and thus of the synchrotron spectrum completely defined by $\nu_{\mathrm{abs}}$, $F_{\mathrm{abs}}$, $\nu_{\mathrm{min}}$, $\nu_{\mathrm{max}}$, and $\nu_{\mathrm{cool}}$.

Except for the high-energy end of the spectrum -- which we impose to cut-off exponentially at $\nu_{\mathrm{max}}$ -- the modeling of the shape of the spectral breaks is taken from \citet[Eq.~(4)]{GS02}. We made, however, the simplification of setting the sharpness of the break to $s\!=\!3/(\beta_1\!-\!\beta_2)$, where the factor $3$ has been chosen to reproduce well the synchrotron self-absorption turnover, which is the most important break. Although this is likely less accurate physically, taking an inverse dependence of $s$ on $(\beta_1\!-\!\beta_2)$ has several advantages. Firstly, it has additive proprieties such that twice the same break at the same frequency  is equivalent to a twice sharper break of $2(\beta_1\!-\!\beta_2)$, thus ensuring smooth transitions when a break crosses the frequency of another break. Secondly, it also works well for negative (concave) breaks and a break of $\beta_1\!-\!\beta_2\!=\!0$ is equivalent to no break, whereas otherwise it changes the normalization of the powerlaw.

The model lightcurve of an outburst is constructed from its evolving spectrum by extracting values at different times, but at a given frequency.
The fitting of a dataset is done simultaneously for all available lightcurves, but only on a subset of all free model parameters at a time. A series of several fits covering all the parameter set, is repeated iteratively until convergence. The number of outbursts and their approximate onset time and amplitude is often defined manually beforehand to guide the iterative fitting process. We implemented the possibility to fit the data with additional constraints to favor solutions with parameter values closer to expectations, e.g. for a simple conical ($r\!=\!1$) and adiabatic ($k_{\mathrm{ad}}$) jet flow with equipartition ($b_{\mathrm{eq}}$) and/or with a minimal dispersion among the specificities $X_{\mathrm{p},i}$ and $\eta_i$ of the different outbursts $i$. Moderate random variations of the uncertainties associated to each data point in a lightcurve before consecutive fit iterations was found to be a good trick to ease convergence in blocked situations, i.e. when the fit is stacked in a local minimum in $\chi^2$, but far from the absolute minimum.

Finally, we note that the contribution from the underlying, undisturbed jet flow can be derived from the physical parameters of the jet by setting $X_{\mathrm{p}}\!=\!0$ and $\eta\!=\!1$, while the cumulative contribution of decaying outbursts peaking before the start of the considered dataset can be obtained by considering a series of outbursts with average values of $X_{\mathrm{p}}$ and $\eta$, spaced by the typical time interval between consecutive events.

\begin{figure*}[t!]
\centerline{\resizebox{0.8\hsize}{!}{\includegraphics[clip=true]{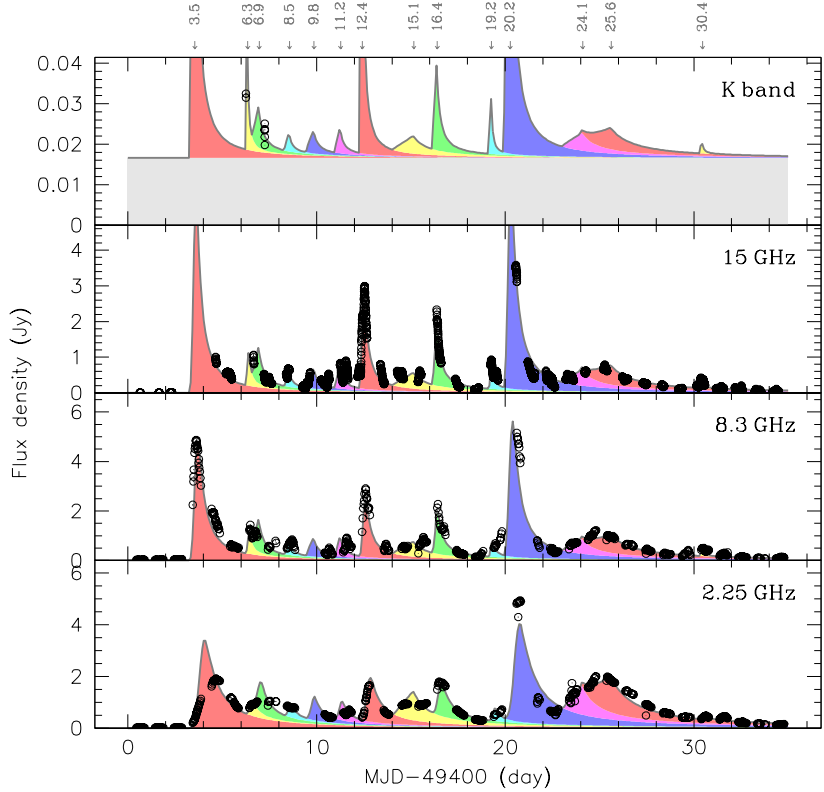}}}
\caption{\footnotesize
Fit of a series of 14 model outbursts -- shown by different colors -- to the flaring episode observed in Cyg X-3 during February--March 1994.
}
\label{turler:f2}
\end{figure*}

\begin{figure}[t!]
\resizebox{\hsize}{!}{\includegraphics[bb=18 144 380 718,clip=true]{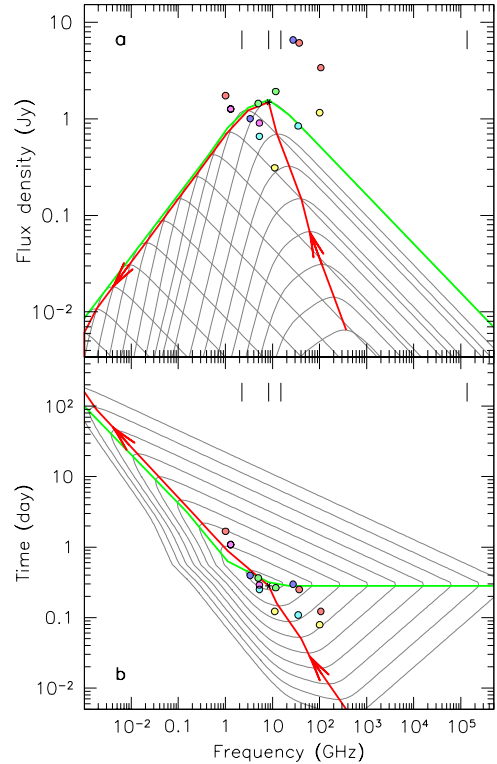}}
\caption{\footnotesize
Evolution of the average model outburst. \textbf{a}) Evolution with time of the peak (red line with arrows) of the synchrotron spectum (thin gray lines) resulting in a peak flux reached at different frequencies (green line). The peak of the overall outburst's evolution is shown by a star symbol, while circles show the position of this peak for each of the 14 individual outbursts, with same colors as in Fig.~\ref{turler:f2}. The four vertical lines at the top indicate the frequencies of the lightcurves in the dataset. \textbf{b}) An areal view on the three-dimensional model outburst in the flux versus time and frequency space. Thin gray lines are contours of equal flux density, whereas the other lines and points are as in the upper panel.
}
\label{turler:f3}
\end{figure}

\section{Results}
\label{turler:results}

As an illustration of the physical modeling described above, we present here a fit to the dataset of Cyg X-3 observed in February--March 1994 \citep{FBW97}. We chose this dataset because it includes a wide variety of outbursts differing in amplitude, time- and frequency-scale, as derived by the more phenomenological approach used by \citet{LTH07}. The resulting fit to the three radio-band lightcurves and the few data points in the infrared K-band is shown in Fig.~\ref{turler:f2}. We obtain a fair description of the dataset with a set of 14 self-similar outbursts, and thus demonstrate that our assumptions of varying only  the strength of the shock and its build-up distance is enough to describe the observed differences among the outbursts. The colored points in the lower panel of Fig.~\ref{turler:f3} show that the trend from short-lived, high-frequency peaking outbursts to long-lasting, low-frequency peaking outbursts is well reproduced by varying the distance, $X_{\mathrm{p}}$, along the jet where the shock compression reaches its maximum. We find an average distance among the outbursts of $X_{\mathrm{p}}\!=\!1.8\times10^{15}$\,cm with typical fluctuations by a factor of 2.5 from one outburst to the next.

We find here that the assumptions of a constant jet flow ($g\!=\!d\!=\!0$) expanding adiabatically ($k_{\mathrm{ad}}$) in a conical ($r\!=\!1$) jet, and with the preferred index $b_{\mathrm{eq}}\!=\!4/3$ for the decrease of the magnetic field was fine for the considered dataset. Because of degeneracy among some of the parameters, we further fixed the value of $f_R$ to $10^{-2}$, $\beta_0\Gamma_0$ to $1$ typical for a moderately relativistic jet with an assumed jet angle to the line-of-sight of $\theta\!=\!10.5\degr$ \citep{MBR04}, and impose an average compression factor of $\eta\!=\!5$, according to $\eta\ltsim 6$ \citep{MG85}. Therefore, the only jet properties that are free to vary are the index, $p$, of the electron energy distribution and the normalizations at $X_0\!=\!10^{16}$\,cm of its low-energy bound, $\gamma_{\mathrm{min},0}$, as well as the corresponding normalizations of the magnetic field $B_0$ and of the electron density $K_0$. The exact value of $\gamma_{\mathrm{max},0}$ was found to be irrelevant and we thus fixed it to $10^5$.

The best fit parameters are an electron index of $p\!=\!2.01$ and a low-energy limit of $\gamma_{\mathrm{min},0}\!=\!4.0$ for the electron energy distribution, with a number density of $K_0\!=\!3.4$\,cm$^{-3}$, and a magnetic field of $B_0\!=\!54$\,mG. These quantities apply to the underlying jet flow at a distance of $X_0\!=\!10^{16}$\,cm from the apex of the jet. For an average outburst peaking at a derived distance of $X_{\mathrm{p}}\!=\!1.8\times10^{15}$\,cm from the apex of the jet, we get peak values of $\gamma_{\mathrm{min,p}}\!=\!21$, $K_{\mathrm{p}}\!=\!1.7\times10^3$\,cm$^{-3}$ and $B_{\mathrm{p}}\!=\!2.5$\,G.
The corresponding energy densities of electrons and magnetic field are $U_{\mathrm{e,p}}\!=\!1.2\times10^{-2}$\,erg\,cm$^{-3}$ and $U_{B,\mathrm{p}}\!=\!0.25$\,erg\,cm$^{-3}$, respectively.

The upper panel of Fig.~\ref{turler:f3} shows the evolution with time of the spectrum of the average outburst. Although this looks very similar to the typical evolution of the shock model of \citet{MG85} without a synchrotron plateau stage (see Fig.~\ref{turler:f1}), the initial rise of the spectrum turnover is of a completely different nature. The parameters we derive do not lead to significant radiative losses (we always have $\nu_{\mathrm{cool}}\gg \nu_{\mathrm{abs}}$), so that we are in the adiabatic cooling stage throughout the outburst evolution. It is mainly the increasing compression $\eta$ and secondly the presence of the low-energy break at $\nu_{\mathrm{min}}$ that define, actually, the initial rise of the spectrum.

\section{Conclusion}
\label{turler:conclusion}

We presented a new, physical modeling and parametrization of outbursts in relativistic jets. We applied the model to a rich dataset of Cyg X-3 and derived the physical conditions in the jet. We found that standard assumptions of a conical, adiabatic jet flow seem adequate and that conditions are such that radiative cooling is negligible. The next step is to fit the model parameters to other datasets as it has been done in the past with a more phenomenological approach. We will then be able to compare physical conditions in the jets of different sources from microquasars to blazars.

An interesting addition to the modeling would be the inclusion of the associated synchrotron self-Compton emission in the X- and gamma-ray spectral domain, where \emph{Swift}/BAT and \emph{Fermi} are currently providing an almost continuous monitoring of bright blazars. This has already been done in a simplified way by \citet{LVT05} and would be particularly interesting for sources where multiple inverse-Compton scattering might play an important role. We would then build on the theoretical study of \citet{B10} to construct a self-consistent description of the multi-frequency emission of shock waves in relativistic jets from the radio to the gamma-rays.

\begin{acknowledgements}
This work was done in the frame of the International Team collaboration number 160 supported by the International Space Science Institute in Bern, Switzerland.
\end{acknowledgements}

\bibliographystyle{aa}

\end{document}